# Design of two combined health recommender systems for tailoring messages in a smoking cessation app

Santiago Hors-Fraile
University of Seville
ETSII Universidad de Sevilla, Avda. Reina Mercedes S/N 41012
Seville, Spain
(+34) 954556817
sanhorfra@gmail.com

Francisco J Núñez Benjumea
Virgen del Rocío University Hospital
Technological Innovation Group
Av Manuel Siurot S/N
(+34) 955013616
francisco.nunez.exts@juntadeandalucia.es

Laura Carrasco Hernández
Virgen del Rocío University Hospital
Smoking Cessation Unit
Av Manuel Siurot S/N
(+34) 635581010
lauracarrascohdez@gmail.com

Francisco Ortega Ruiz
Virgen del Rocío University Hospital
Smoking Cessation Unit
Av Manuel Siurot S/N
(+34) 955013163
francisco.ortega.sspa@juntadeandalucia.es

Luis Fernandez-Luque
Salumedia Tecnologías
Sevilla, Spain
(+34) 656930901
luis@salumedia.com

## ABSTRACT
In this article, we describe the design of two recommender systems (RS) designed to support the smoking cessation process through a mobile application. We plan to use a hybrid RS (content-based, utility-based, and demographic filtering) to tailor health recommendation messages, and a content-based RS to schedule a timely delivery of the message. We also define metrics that we will use to assess their performance, helping people quit smoking when we run the pilot.

## 1. INTRODUCTION
Smoking is responsible for severe conditions such as exacerbation of asthma, pulmonary fibrosis, chronic obstructive pulmonary disease (COPD), and lung cancer, among others [1]. People who try to quit suffer from nicotine abstinence syndrome. They usually experience cravings, headaches, intestinal disorders, weight gain, insomnia, dullness and irritability among other withdrawal symptoms [2]. Although tobacco prevalence differs substantially among countries [3], statistics show how people still try to quit [4] and that the success rate has room for improvement.

The high penetration of smartphones in everyday lives lets us reach people in a way that we could not before [5]. Mobile devices, and their associated mobile application software – referred from now on as apps – have brought a new channel of communication. Since they are ubiquitous, we can reach patients almost anywhere anytime. Besides, smartphones let us have a real-time feedback loop to understand what, how, and when patient support is most effective and well-received.

In order to help people who want to quit smoking, many authors have shown that apps are a suitable tool [6]. Despite their results, there is still a need to find out how patients can receive more effective support. Several authors have proved that personalized and tailored apps may increase the effectivity of the lifestyle recommendations that patients can get via their smartphones [7].

In this context, one of the things the SmokeFreeBrain (SFB) project – funded by the EU within the Horizon 2020 program – aims to explore is how effective an app can be throughout the smoking cessation process.

Among other features, the app prompts users motivational messages after the quitting date. The message frequency is based on the trans-theoretical Behavior Model [8], as some studies have already proposed [9]. In order to maximize the effectiveness of the system, the following questions should be addressed: What type of topic motivates each user more? When does each user prefer to receive the message so that the message is more helpful and not intrusive?

To solve these questions, we included two RSs in a server that worked in coordination with the app. One of them was a hybrid RS to select the messages of those topics more relevant for the user, that actually help them change their behavior. The other RS was to tune the time in which those messages were more effective. Each night a scheduled task runs to calculate when and which message has to be sent for each user. We will measure the impact of the RS outcomes both with the in-app metric statistics and a satisfaction questionnaire at the end of the pilot.

## 2. METHODS
### 2.1 Pilot protocol overview
This is a randomized open-label parallel-group trial. For eight months, all patients attending the SCU of the VRUH will be offered to join this study. The inclusion criteria are: Smoking population attending to the SCU of VRUH, subjects older than 18 who want to give up smoking with Android-based smartphones and the ability to interact with the smartphone; subjects who can also sign an Informed Consent Form. Subjects who have adverse effects related to the pharmacological treatment included in the study are to be excluded. No economic compensation is provided to participate.

All patients in the study will follow the usual care (psychopharmacological treatment) provided at the SCU of the VRUH for one year – unless they drop out. This treatment consists of a first interview of the patient with a psychologist and a pulmonologist in order to establish familiarity among the patient and the healthcare professionals, know the patient's smoking habits and aims, and to set a quitting date with a quitting plan when healthcare professionals consider the patient ready. A fortnight after the quitting date, a new meeting is held to follow-up on the patient. Until the end of the one year tobacco abstinence date – condition to which a patient is discharged – several follow-up consultations are held either face-to-face or by telephone. These consultations are set at approximately 15, 30, 60, 90, 120, 180, 365 days from the base consultation. In such consultations, dependency, motivation and quality of life are assessed, as well as exhaled CO and urine cotinine levels.

All patients will be provided with free drugs prescribed by healthcare professionals – varenicline and bupropion. However, patients in the intervention group are provided with a complementary app that contains information about smoking cessation, offers personal physical activity level (PPAL) comparison information, and has some relaxing and distracting tools to help patients pass cravings for cigarettes (for instance, mini-games). This app includes a motivational message notification system.

## 2.2 Message type selection recommender system algorithm

Every night, a server computes what motivational messages should be received by each patient the next day. The motivational messages that are sent to patients via the smartphone can be taken from five different pools: general motivation, diet tips, exercise and active life recommendations, PPAL, and tobacco smoking consequences. The first three groups are easy to understand by just reading their title. The fourth pool is related to each patient's PPAL. We generate these messages using the Google Fit activity data, for instance "Hello Peter! You did great yesterday! You were 15 min over your average activity time. Keep up the good work today!" The fifth pool has information related to the negative consequences of smoking tobacco, for instance, "People who smoke develop Reinke's edema more easily – a liquid retention in their necks. It's great that you don't smoke anymore! Kind regards, Dr. Laura Carrasco". Each pool contains 150 different messages dealing with the topics under the pool titles.

Since patients get a limited number of messages – flooding them with messages may trigger their desire to smoke even more [9] – we have to optimize the type of messages they like. We faced this problem with a hybrid RS. It has a parallelized hybridization design. Following Burke's taxonomy [10] we can state that this RS follows a weighted strategy. It combines three individual algorithms: a utility based algorithm, a demographic algorithm, and a content-based algorithm.

Given that a user is "u", an item "i", and a weighting function "β", and that "k" individual algorithms are forming the hybrid algorithm, the general formula for the recommendation of an item "i" for a user "u" is:

$$rec_{weighted}(u,i) = \sum_{k=1}^{n} \beta_k \times rec_k(u,i) \quad [13]$$

We wanted that the β weights were different for each individual algorithm. Specifically, the more number of neighbors associated to a user, and the more number of ratings that one user gave, the less value the content-based algorithm had. We did this due to the nature of the user base in our system: a relatively small number of users (up to 120), a gradual user sign-up that would increase the cold start problem, and a gradual user drop-out as they finished their one year treatment. Thus, we want that the content based algorithm's importance diminishes as the number of users and ratings increase. We used the followed functions to calculate each weight for a given user "u"

$$\beta_{demographic}(u) = 1 - \frac{50}{n_u^2 + 50}$$

$$\beta_{utility}(u) = 1 - \frac{50}{r_u^2 + 50}$$

$$\beta_{utility}(u) = 1 - \frac{\beta_{demographic}(u) + \beta_{utility}(u)}{2}$$

where $n_u$ represents the number of neighbors for a given user "u" in the recommender system, and $r_u$ represents the number of ratings a given user "u" has performed. The graphical representation of the weight function for the demographic filtering algorithm (Figure 1) shows that for values higher than 0, we always get values between 0 and 1. For values higher than 22, we get a result higher than 0.9. This might even be considered a switching function for large values of the number of neighbors. This behavior was intentionally sought because as Desroisers and Karypis suggested [11], collaborative filtering starts producing fairly accurate results when it considers between 20 and 50 neighbors. Since our number of users would be small, we would go for the lower boundary of 20.

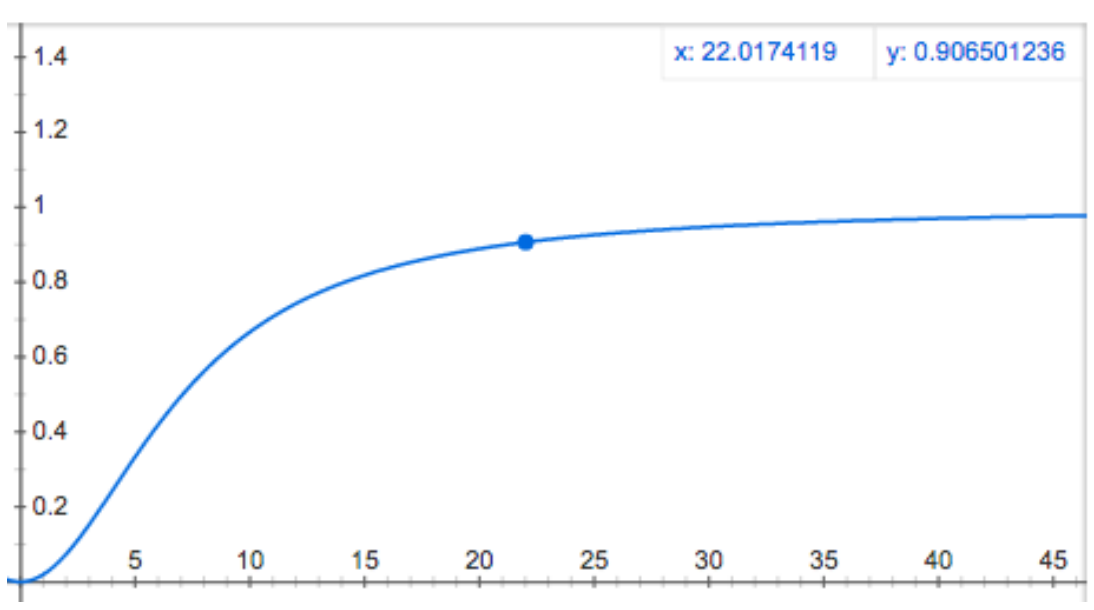

**Figure 1: Graphical representation of the empirical weight formula 1-(50/(x²+50))**

### *2.2.1 Demographic filtering algorithm*

The demographic filtering algorithm we use was inspired in Ge's work [12]. However, we designed ours as a collaborative filtering algorithm in which the similarity is the mean average of the other users' message ratings, and the following users' attributes: gender, employment status, age, quitting date, Fagerström test and the Richmond test score. Since we will have a reduced number of users – a maximum of 120 users - we consider all users in the system.

To calculate the similarity between users "$u_1$" and "$u_2$" based on their ratings we use the Pearson's formula. To calculate their similarity based on their attributes we compute the mean average of the absolute value of the difference between their attribute values as follows:

$$similarity(a_{u_1,}a_{u_2}) = \frac{1}{a_{u_1} - a_{u_2}}$$

For the special cases of non-numeric attributes – that is, the gender, and employment status - we set a "0" value if they are not equal, and a "1" value if they are equal.

### *2.2.2 Utility-based algorithm*

For the utility-based algorithm we use both explicit and implicit utility rates. The explicit ratings are retrieved from the direct ratings users give to each message they receive. Users can rate a message whenever they open the message. They can vote for the message with a "like", "dislike" or "neutral" button. Each read message has to be rated at least once, although users will be able to change their votes in the future by opening a message and rating it again. The new votes override the previous ones.

The implicit ratings are calculated using the interaction activity of the user with different parts of the app.

**Table 1. User actions that influence the implicit utility**

| User actions | Affected message type | Constraints |
|---|---|---|
| User accesses a General Motivation content section. | General Motivation | User has to stay at least 4 seconds in the section |
| User reads a General Motivation message. | General Motivation | None |
| User accesses a Diet tips content section. | Diet tips | User has to stay at least 4 seconds in the section |
| User reads a Diet tip message. | Diet tips | None |
| User accesses an Exercise and active life tips content section. | Exercise and active life tips | User has to stay at least 4 seconds in the section |
| User reads an Exercise and active life tip message. | Exercise and active life tips | None |
| User accesses the PPAL section. | PPAL | User has to stay at least 4 seconds in the section |
| User reads a PPAL message. | PPAL | None |
| User accesses a Smoking consequences content section. | Smoking consequences | User has to stay at least 4 seconds in the section |
| User reads a Smoking consequence message. | Smoking consequences | None |

The final implicit rating is the average of two elements. The first one is the quotient between the user's number of accesses to each section, and the total number of accesses to all sections. The second element is the quotient of the number of accesses to each message type, and the total number of accesses to all messages.

Then, we compute the probability of each message type "t" for a given user "u" for normalized probability as follows:

$$P(t,u) = \left(1 - \frac{1}{v_u}\right) \times e_t + \left(\frac{1}{v_u}\right) \times i_t$$

where $v_u$ is the total number of votes of the user "u", $e_t$ is the explicit rate user "u" provided for the message type "t", and $i_t$ is the implicit rate the user "u" provided for the message type "t".

#### *2.2.3 Content-based algorithm*

The content-based algorithm is based on an interest list defined by each user. Each user fills out a form when they open the app for the first time and express their topics of interest in a 3-level Likert-type scale. They are able to change them later. The possible interests are the five different existing message types explained in section 2.2. Thus, we calculate the similarity as the cosine similarity between the interests' vectors the user directly generates with the form, and the unit vectors for each of the five 5-element vector.

### 2.3 Message sending time recommender system algorithm

In order to send the users timely motivational messages, we designed an utility-based algorithm with some constraints. The aim of this recommender system is to avoid the "robotic" automatic message feeling that patients may have when they get their message always at the same time [9]. Nevertheless, we also wanted to avoid an erratic and bothersome random message sending time.

We solved this by requesting patients to set a daily time frame that defines when they would like to get messages the first time they use the app. For instance, from 8:00 until 21:00. Thus, the messages for that user will only be sent between 8:00 and 21:00. The minimum time frame that can be set is of one minute, which will virtually turn this into an almost fixed-time message sender. Patients may opt to leave the time frame set to the default 24 hours.

We define the utility of each message as the time between the delivery and reading. The shorter the time, the more the utility the chosen delivery time has. Our hypothesis is that if a patient reads a message as it is sent, it is because she is receptive and in a good moment of her day to get that tip.

Once the message sending time is chosen, the server is notified so that it sends the notification at the specified time.

### 2.4 Selection

The output for each of the two previously explained RSs is a vector with the possible options to choose from (O), and their corresponding probabilities (P). Instead of just picking the option with highest probability, we use a probability function based on P which will determine O. That is, we compute a random number between one and the sum of all P components. Then, we check what component the random number belongs to, and we select the option in O that corresponds to the component in which the random number matched. We can see a general view of the SFB message recommendation process in Figure 2.

## 3. EVALUATION DESIGN

The evaluation setting for these RSs are constrained. There are no natural or synthetic datasets that we can use to test their efficacy. The standard metrics (MAE, hit rate, F1, rank score, etc.) are not applicable to our case since our RS selects a message and sends it at a specific time. This prevents the user to interact with other candidate items. Besides, we will not be able to have them evaluated with A/B testing due to the constraints of the pilot – the intervention group has to have the same app. Comparing the quitting rates between the intervention and control group would also be inaccurate because there are more variables in the app such as the minigames to pass cravings, and the contents of the app that may be affecting the smoking cessation rates. That is, we would be measuring the whole e-health intervention, not just the proposed RS.

Thus, we propose a nonexperimental user study in which we measure the rates of correct predictions and false positives as the main method to rate the accuracy of our RSs. We will not be able to assess the false negatives or correct omissions since in our case users will only see and rate what the RSs give them.

In addition, we will include two 7-option Likert-type RS-specific questions in a post-trial survey that all pilot participants of the intervention group will have to take after they complete the one-year pilot. These questions will be translated to Spanish, the native language of our users:

- How did you find the message topic tailored to your preferences?
- How do you rate the time the messages were sent?

Regarding the performance in terms of the execution time of the RSs, they will not be run on-demand by the users, but run during the night, and then the results sent during the following days. Thus, we consider unnecessary for our aims to assess the processing speed of the computations.

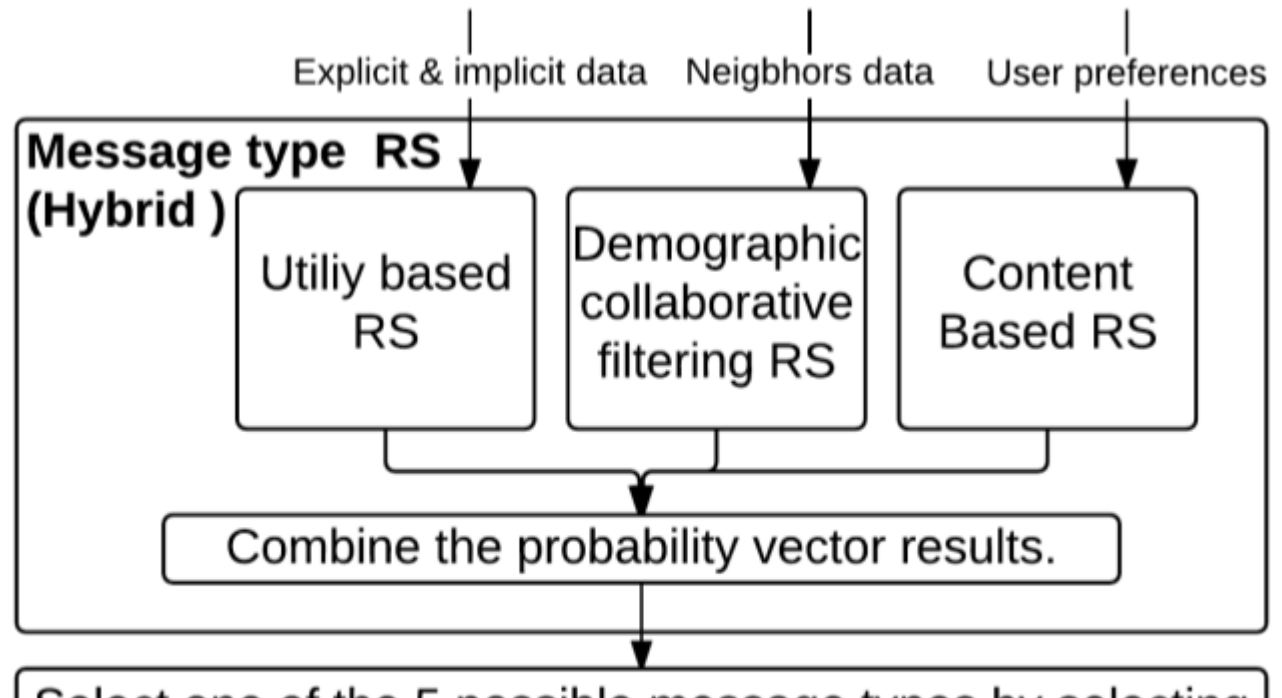

## 4. DISCUSSION

The app developed in the frame of the SFB project introduces a combination of two RSs to tailor messages that will be delivered through a smoking cessation app. From a total of 240 who will be recruited in this pilot, 120 patients will be part of the intervention group.

We hope the reduced complexity of the presented recommendation items compared with other larger systems – five types of messages (items) and six potential time frames – facilitate us in making a deeper data analysis. The evaluation design was planned bearing in mind the need for other researchers to understand better how RS can be used for timely recommendations about lifestyle.

Although it is out of the scope of this study, future research is encouraged to test other weighting formulas to the hybrid algorithm, and also for the demographic similarity function – including the attributes. Also in our implementation of the message sending time, users may not add multiple time frames for the same day. Similarly, they cannot add different time frames for different days of the week. Further research to explore these possibilities of having more flexibility to define the patients' preferred message receiving time is desired to see whether it has a relevant impact.

## 5. ACKNOWLEDGEMENTS

The present study was funded through the Project SmokeFreeBrain "Multidisciplinary tools for improving the efficacy of public prevention measures against smoking" of the European Union's Horizon 2020 research and innovation programme under the grant agreement No 681120.